\documentclass[preprint,showpacs,preprintnumbers,amsmath,amssymb]{revtex4}

\usepackage{graphicx}
\usepackage{dcolumn}
\usepackage{bm}

\begin{document}

\title{\large  Strongly suppressed proximity effect and ferromagnetism in topological insulator / ferromagnet / superconductor thin film trilayers of
$\rm Bi_2Se_3$ / $\rm SrRuO_3$ / underdoped $\rm YBa_2Cu_3O_x$: \\
A possible new platform for Majorana nano-electronics }

\author{Gad Koren}
\email{gkoren@physics.technion.ac.il} \affiliation{Physics
Department, Technion - Israel Institute of Technology Haifa,
32000, ISRAEL} \homepage{http://physics.technion.ac.il/~gkoren}

\date{\today}
\def\bfig {\begin{figure}[tbhp] \centering}
\def\efig {\end{figure}}

\normalsize \baselineskip=8mm  \vspace{15mm}


\pacs{74.78.Fk, 74.25.Sv, 75.70.-i, 73.20.-r }

\begin{abstract}

We report properties of topological insulator - ferromagnet - superconductor trilayers comprised of thin films of 20 nm thick $\rm Bi_2Se_3$ on 10 nm $\rm SrRuO_3$ on 30 nm $\rm YBa_2Cu_3O_x$. As deposited trilayers are underdoped and have a superconductive transition with $\rm T_c$ onset at 75 K, zero resistance at 65 K, $\rm T_{Cueri}$ at 150 K and $\rm T^*$ of about 200 K. Further reannealing under vacuum yields the 60 K phase of $\rm YBa_2Cu_3O_x$ which still has zero resistance below about 40 K. Only when $10\times 100$ micro-bridges were patterned in the trilayer, some of the bridges showed resistive behavior all the way down to low temperatures. Magnetoresistance versus temperature of the superconductive ones showed the typical peak due to flux flow against pinning below $\rm T_c$, while the resistive ones showed only the broad leading edge of such a peak. All this indicates clearly weak-link superconductivity in the resistive bridges between superconductive $\rm YBa_2Cu_3O_x$ grains via the topological and ferromagnetic cap layers. Comparing our results to those of a reference trilayer with the topological $\rm Bi_2Se_3$ layer substituted by a non-superconducting highly overdoped $\rm La_{1.65}Sr_{0.35}CuO_4$, indicates that the superconductive proximity effect as well as ferromagnetism in the topological trilayer are actually strongly suppressed compared to the non-topological reference trilayer. This strong suppression is likely to originate in strong proximity induced edge currents in the SRO/YBCO layer that can lead to Majorana bound states, a possible signature of which is observed in the present study as zero bias conductance peaks. \\


\noindent Keywords: superconductor, topological insulator, ferromagnet, Thin film trilayer\\

\end{abstract}

\maketitle

\section{Introduction}
\normalsize \baselineskip=6mm  \vspace{6mm}

Thin film junctions, bilayers and trilayers of a topological insulator (TI), a ferromagnet (FM) and a superconductor (SC), can exhibit many interesting phenomena such as triplet superconductivity \cite{KorenTriplet,FuBerg}, topological superconductivity \cite{FuBerg,KaneRMP,Kitaev}, various proximity effects \cite{Koren1,Koren2,YangAndo,WeiMoodera,KorenMPE} and lead to the creation of exotic states including Majorana fermions \cite{Kouwenhoven,Moti,CMarcus}. All these phenomena involve strongly correlated electrons and the challenge is to understand the mutual effects of the interactions between the different comprising layers. Following our previous investigation on TI/FM bilayers of $\rm Bi_{0.5}Sb_{1.5}Te_3$ / $\rm SrRuO_3$  (BST/SRO) where a robust magnetic proximity effect with the seemingly weak itinerant FM was found \cite{KorenMPE}, we decided to extend our study to TI/FM/SC trilayers with the goal of creating spatially separated pairs of Majorana bound states (MBSs) in topological superconductors (TSCs). In the presence of the FM layer, time reversal symmetry (TRS) is broken, thus in principle MBS in TSC in 2D can be formed in these trilayers \cite{Beenakker,Ramon}. In the present paper we make an attempt to realize a 2D  network of in-plane SC-TI/FM-SC junctions in TI/FM/SC thin film trilayers. A theoretical study on a structure similar to our trilayers, including all the ingredients of a topological phase, a ferromagnet and SIS Josephson junctions has been published recently and seems to be relevant to the present results \cite{WuLevin}. Furthermore, effort to fabricate networks comprised of 1D SC/TI nano-wires hosting MBSs was also recently reported \cite{Marcus}. These networks should allow for scaling up of the number of such devices (the MBSs at the ends of the wires) on a wafer, braiding them, and may open the way for Majorana nano-electronics for quantum computing.  \\

When measuring transport properties with a bias current parallel to the interface of multilayers which contain a superconducting layer, the problem is always shorting by the SC layer. This results from the zero resistance of the SC layer below $\rm T_c$ which carries all the current without observable influence of the other layers. To avoid this shorting effect, very thin SC layers should be used, or even better, layers of SC islands. In the cuprate this can be achieved by underdoping which depletes oxygen mostly from the inter-grain regions and rendering them non-superconducting. This was done in the present study, where TI/FM/SC trilayers of the topological insulator $\rm Bi_2Se_3$ (BS), the itinerant ferromagnet $\rm SrRuO_3$ (SRO) and the d-wave superconductor $\rm YBa_2Cu_3O_x$ (YBCO) were used, with a ferromagnetic and superconducting transition temperatures of $\rm T_{Curie}=150$ K and $\rm T_{c}=60$ K, respectively. Weak-link islands (or grains) superconductivity was found, where the coupling between the SC islands is via the TI/FM cap layers, forming a network of 1D current carrying channels. By comparing the results to those of a reference trilayer where the TI layer is replaced by highly overdoped $\rm La_{1.65}Sr_{0.35}CuO_4$ (LSCO35), we find that the superconductive proximity effect and ferromagnetism in the topological trilayer are in fact strongly suppressed as compared to those of the non-topological reference trilayer. We attribute this suppression to proximity induced edge currents in the FM/SC bilayer by the TI at the interface, which can host MBSs in the weak-link 1D channels.\\

\section{Preparation and basic properties of the films\\
and trilayers }
\normalsize \baselineskip=6mm  \vspace{6mm}

Thin films of YBCO and SRO were chosen as the superconductor and ferromagnet for the present study since both are compatible and lattice matched with one another and grow epitaxially on (100) $\rm SrTiO_3$ (STO), without inter diffusion between the layers. They are also very well characterized in the literature \cite{Koren89,Marshall,Klein}. $\rm Bi_2Se_3$ (BS) is the most well known topological insulator (TI) and its thin films are also well characterized \cite{Zhang,Butch}. These were used to prepare and investigate the topological trilayers (TTL) of BS/SRO/YBCO. For comparison, a reference trilayer (RTL) was prepared without a topological layer. LSCO35 was chosen to substitute for the BS layer in the trilayer (yielding LSCO35/SRO/YBCO) since it is compatible with the base SRO/YBCO bilayer, grows epitaxially on it under the same deposition conditions, is non-superconducting and has a similar carrier concentration to that of the BS layer. All films were prepared on (100) STO wafers of $10\times 10$ mm$^2$ area by pulsed laser deposition using the third-harmonic of a Nd-YAG laser. The laser was operated at a pulse rate of 3.33 Hz, with fluence of $\rm \sim 1.5\, J/cm^2$ on the targets for the deposition of the epitaxial YBCO, SRO  and reference LSCO35 films. Lower fluence of $\rm \sim 0.6\, J/cm^2$ was used for the deposition of the BS layer. The SRO/YBCO bilayer as well as the LSCO35/SRO/YBCO trilayer were prepared \textit{in-situ} at 800 $^0$C in one deposition run, while the BS film was deposited on the SRO/YBCO bilayer at 250 $^0$C in a different vacuum chamber. The YBCO, SRO and LSCO35 films were deposited using stoichiometric ceramic targets of $\rm YBa_2Cu_3O_x$, $\rm SrRuO_3$ and $\rm La_{1.65}Sr_{0.35}CuO_4$ under 70-90 mTorr of $\rm O_2$ gas flow at 965 $^0$C heater block temperature (about 800 $^0$C on the surface of the wafer). The $\rm Bi_2Se_3$ layers were deposited under vacuum using a pressed target with Bi to Se ratio of 1:17 at 300 $^0$C heater block temperature (about 250 $^0$C on the surface of the wafer). Electron dispersive spectroscopy (EDS) measurements showed that disregarding the stoichiometry of the target, the resulting films had a Bi to Se ratio of 2:3 which is the stable Bi-Se phase. X-ray diffraction measurements showed that this film had the hexagonal $\rm Bi_2Se_3$ structure, with preferential c-axis orientation normal to the wafer and c=2.84 nm.\\

First annealing of the SRO/YBCO bilayer and LSCO35/SRO/YBCO trilayer was done \textit{in-situ} under 0.5 atm. of O$_2$ pressure with a dwell time of 1 h at 400 $^0$C. This yielded well oxygenated samples with T$_c$ of 88 K. Deposition of the BS layer on the SRO/YBCO bilayer was performed under vacuum at 250 $^0$C and took about 20 minutes, after which the sample was rapidly cooled down  to room temperature. This yielded the "as deposited" TTL which was underdoped with T$_c$ of 75 K, $\rm T_{Curie}$ of 150 K (SRO is insensitive to this vacuum annealing) and T$^*$ of the pseudogap of underdoped YBCO at about 200 K. Further depletion of oxygen to obtain the 60 K YBCO phase, was done by annealing under vacuum at 250 $^0$C for two more hours. This second annealing step was applied also to the RTL of LSCO35/SRO/YBCO. Patterning of the trilayers was obtained by Ar-ion milling at 13 mA/cm$^2$ and -180 $^0$C, to minimize damage to the samples by heating during the milling process. Generally, ten equidistant microbridges of $\rm 10\times 100\,\mu m^2$ were patterned across the line dividing the wafer into two halves, with 1 mm distance between them. Forty $\rm 1\times 1\,mm^2$ contact pads were patterned on the two sides of the dividing line, 20 on each side. Transport measurements were carried out by the use of an array of 40 gold coated spring loaded spherical tips for the 4-probe dc measurements on ten different locations on the wafer, either with or without silver paste or gold contact pads.  Magnetic fields of up to 8 T were applied normal to the wafer along the c-axis direction of the YBCO, BS and LSCO35 layers.

\section{Results and discussion}

\begin{figure} \hspace{-20mm}
\includegraphics[height=8cm,width=10cm]{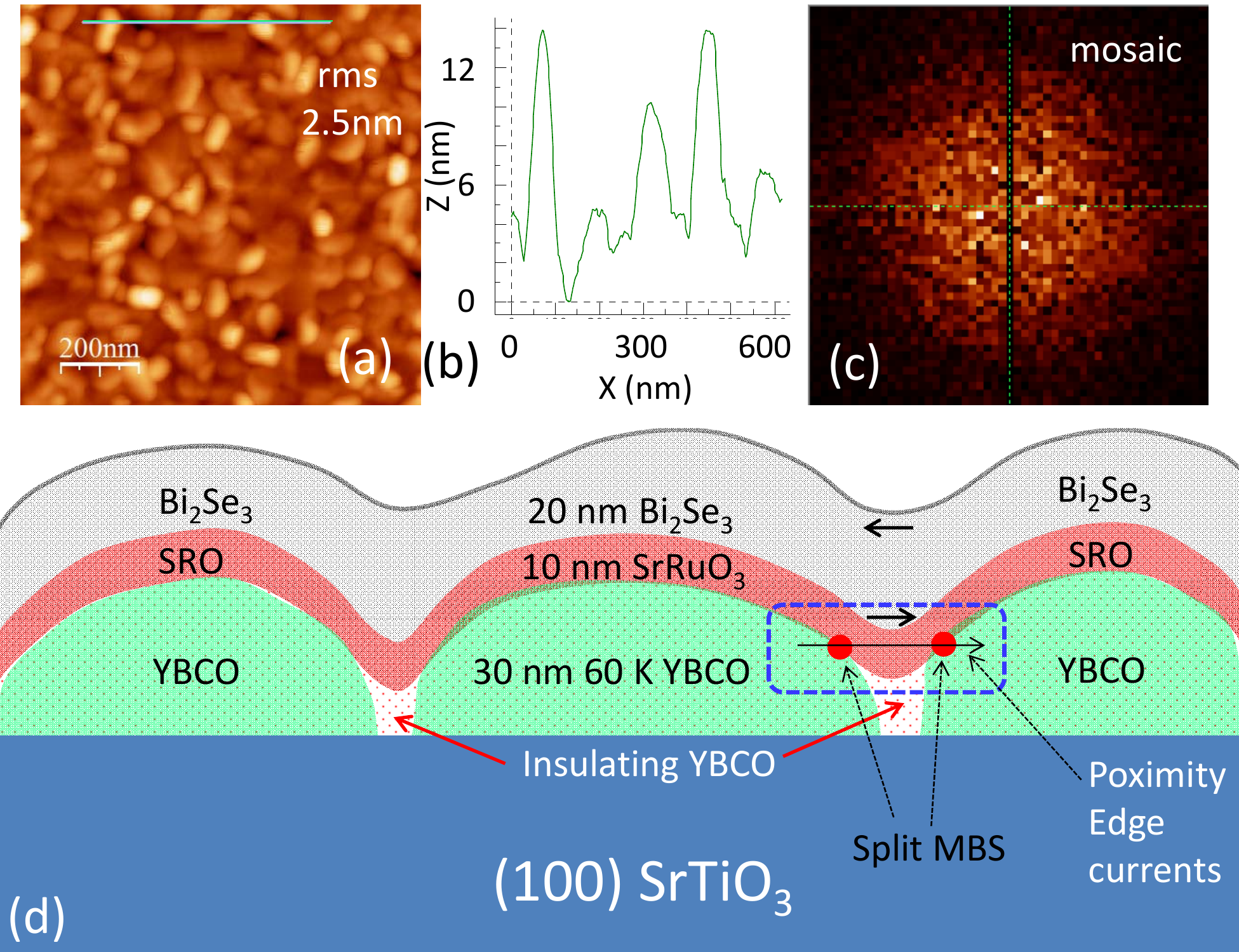}
\hspace{-20mm} \caption{\label{fig:epsart} (a) Atomic force microscope image of the 20 nm BS on 10 nm SRO on 30 nm YBCO trilayer. Height profile along the line in (a) is shown in (b). Fourier transform of (a) is given in (c). (d) depicts a schematic trilayer cross section model with three weakly linked YBCO islands. The enclosed area in the dashed rectangle shows a proposed scenario for the creation of split MBS similar to that described by Beenakker  \cite{Beenakker}.      }
\end{figure}

We begin with characterizing the surface morphology, crystallization and in-plane order of the crystallites  in the presently used TTL (20 nm BS/10 nm SRO/30 nm YBCO). Fig. 1 (a) shows an atomic force microscope image of a $\rm 1\times 1\,\mu m^2$ area of this TTL, while in (b) a typical height profile is shown along the line in (a). One can see good crystallization with rms roughness of 2.5 nm, which compared to the overall thickness of the TTL (60 nm), yields 4\% roughness. Thus the TTL is quite smooth. The Fourier transform of (a) is given in (c) which reveals quite uniform angular intensity distribution in k-space, indicating mosaic in-plane order of the crystallites in real space. This is a result of a few different hexagonal in-plane orientations (see the bright spots in (c)) but without long range order. The fact that the BS layer is hexagonal here was established by the x-ray diffraction data where the hexagonal c-axis value was found (2.84 nm). Fig. 1 (d) depicts a schematic model for the underdoped TTL where three 60 K YBCO islands (or grains) are separated by narrow insulating YBCO regions (the white areas with red dots). These islands are nevertheless weakly connected via the BS/SRO cap layer, which leads to weak-link superconductivity between the grains via the proximity effect. Our results will be discussed in view of this model, and as we shall see they can be described well by it.\\

\begin{figure} \hspace{-20mm}
\includegraphics[height=8cm,width=10cm]{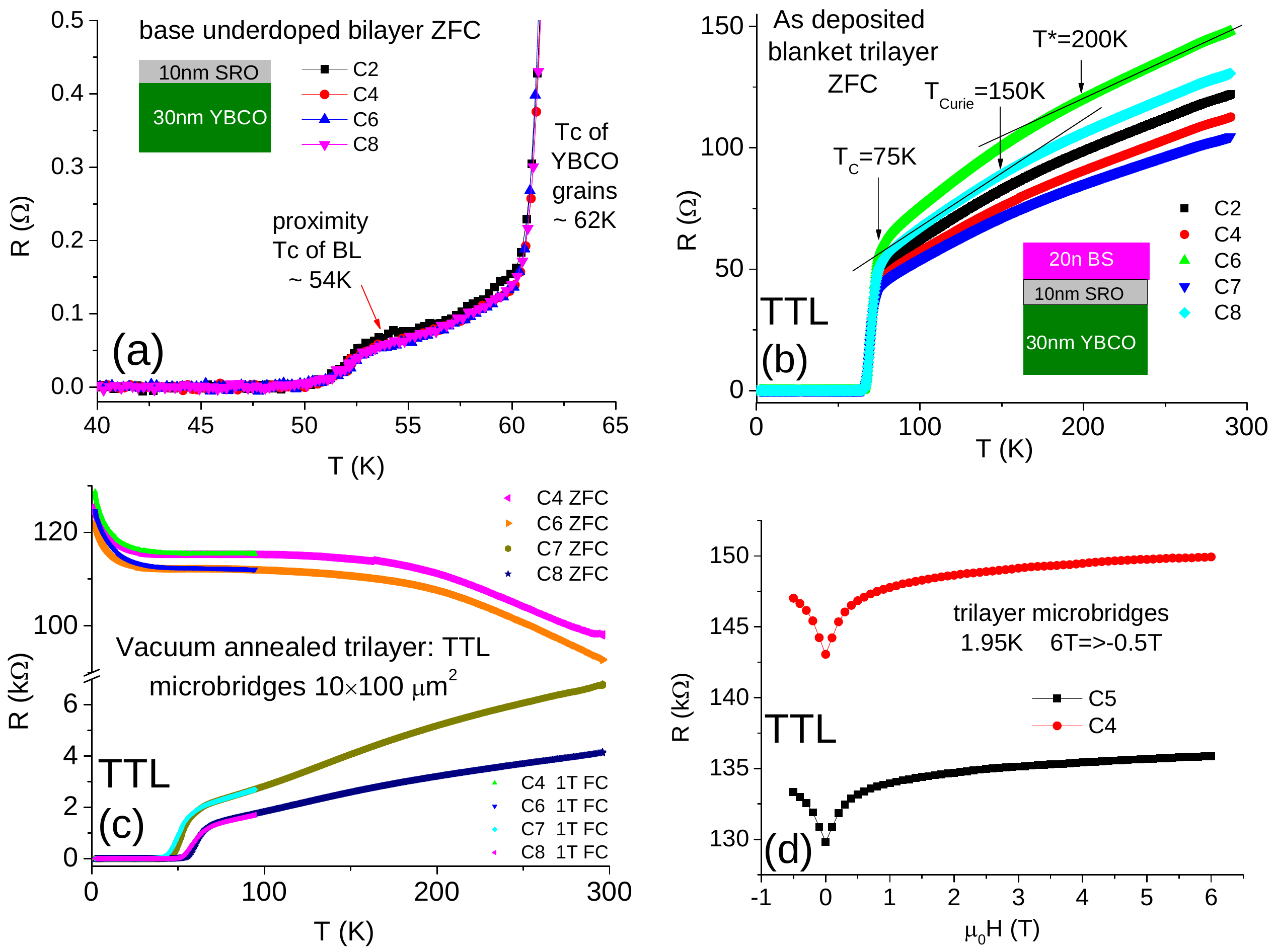}
\hspace{-20mm} \caption{\label{fig:epsart} Resistance vs temperature of a blanket base bilayer of 10 nm SRO on 30 nm YBCO common to all trilayers in this paper is shown in (a). (b) shows R vs T of the as-deposited TTL of 20 nm thick BS on 10 nm SRO on 30 nm YBCO before patterning, while (c) shows it after patterning into ten $\rm 10\times 100\,\mu m^2$ microbridges and further vacuum annealing. (a), (b) and (c) were measured  under ZFC, and (c) also includes 1 T FC. (d) depicts R vs magnetic field H of two resistive TTL microbridges of (c) at 1.95 K.   }
\end{figure}

Fig. 2 (a) depicts R vs T results under zero field cooling (ZFC) measured at four different locations on a base bilayer of 10 nm SRO/30 nm YBCO which is common to all trilayers in the present paper. The main superconducting transition temperature of the underdoped YBCO layer is about 62 K. Below it, a second small proximity effect (PE) transition with $\rm T_c\sim 54$ K can be seen (the knee in this curve). If the YBCO grains in this bilayer would have formed a continuous superconducting layer, then the bias current would flow through it masking any PE with the cap SRO layer. Thus the YBCO grains are separated by narrow insulating regions as depicted in Fig. 1 (d), and the PE via the SRO cap layer is established. Fig. 2 (b) shows the as deposited R vs T results of the 20 nm BS/10 nm SRO/30 nm YBCO TTL which is underdoped with $\rm T_c$ of 75 K. The reason for the underdoping is that the topological BS layer is deposited at 250 $^0$C under vacuum, and already at this temperature the YBCO layer loses oxygen, more so in the grain boundary regions (see Fig. 1 (d) again). Nevertheless, the YBCO grains here are still connected too strongly, shorting the TTL below $\rm T_c$(R=0)  = 62 K, with a small PE "knees" above it up to 65 K. Only after further two hours vacuum annealing at 250 $^0$C, this TTL had $\rm T_c$(R=0) of 50-53 K. Fig. 2 (c) shows R versus T results of four underdoped $\rm 10\times 100\,\mu m^2$ microbridges patterned on this TTL. One set of data depicts ZFC results from room temperature down to 2 K, while the other represents 1 T field cooling (FC) data from 95 K down to 2 K. A striking result is that while the unpatterned sample shows transition to superconductivity all over the wafer with onset at 68 K, the microbridges show completely different results. Two of the bridges show a transition to superconductivity at 67 and 62 K with about 1 $\mu \Omega$ below 53 and 45 K, respectively, whereas the other two are much more resistive at low temperatures and show insulating behavior versus T (note the broken ordinate scale in Fig. 2 (c)). Clearly, the different locations of the bridges on the wafer sample different small areas of the trilayer, which either exhibit percolative superconductivity (as in C7 and C8) or not (as in C4 and C6). Fig. 2 (d) depicts R vs magnetic field H of two resistive TTL microbridges of (c) at 1.95 K. The result is very similar to the magnetoresistance of a stand-alone BS topological layer. It shows weak anti-localization (WAL) at low fields up to about 1 T with tendency to saturation above it \cite{KorenMR,HLN}. Unlike previous magnetoresistance results in bilayers of 15 nm BST/15 nm SRO where the WAL signature of the TI was observed together with that of the coercive field of the FM layer \cite{KorenMPE}, here no effect of either the FM or SC layers is observed. This could be due to the fact that only a thinner FM layer of 10 nm SRO is used in the present study, but as we shall see later this is not the case, and both the superconductive PE and ferromagnetism are strongly suppressed in the present TTL.\\

\begin{figure} \hspace{-20mm}
\includegraphics[height=8cm,width=10cm]{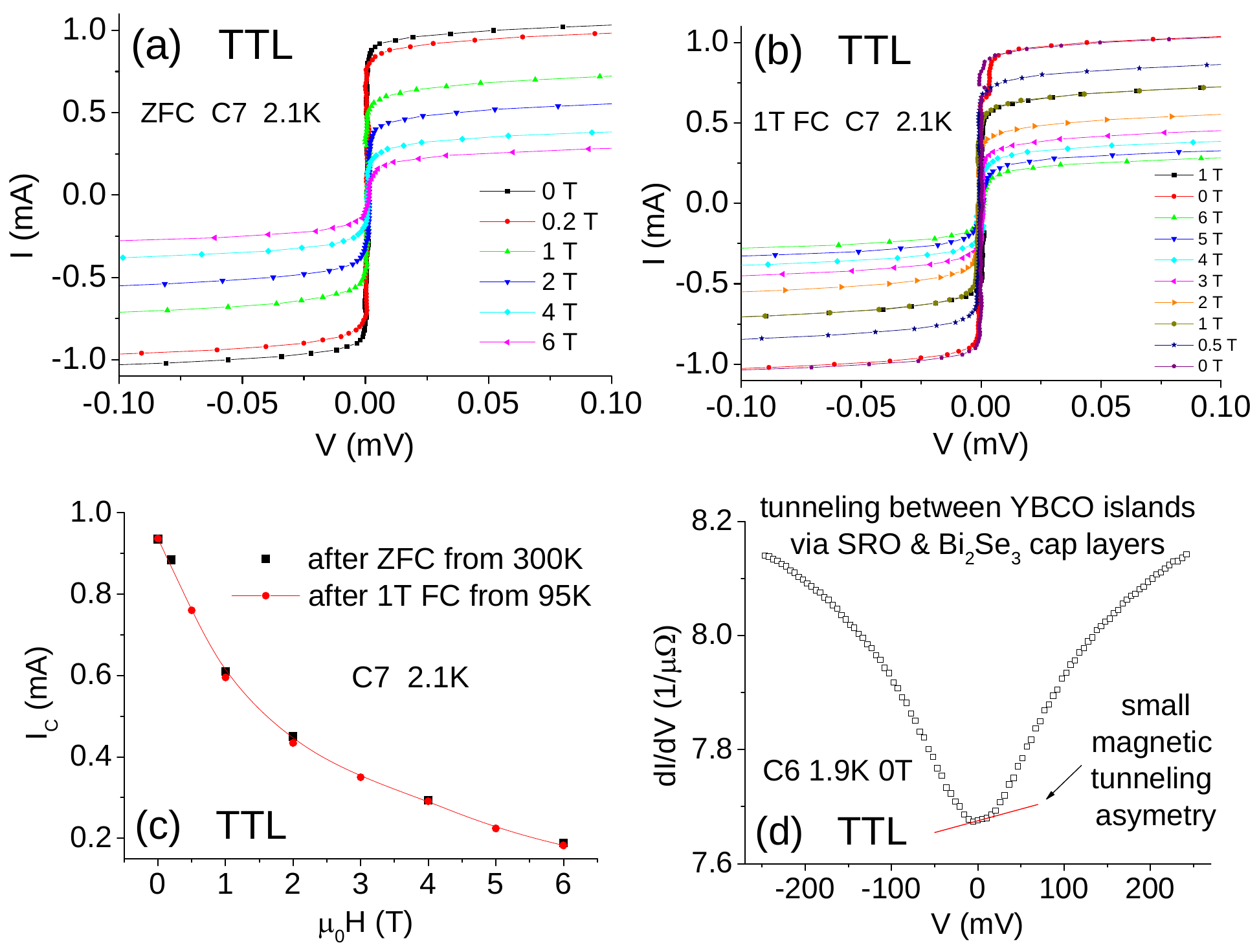}
\hspace{-20mm} \caption{\label{fig:epsart} I-V curves of the C7 microbridge of the TTL of Fig. 2 (c) at 2.1 K under different magnetic fields after two cooling conditions: ZFC from room temperature in (a) and 1 T FC from 95 K in (b). (c) shows the critical currents $\rm I_c$ at 2.1 K versus field as deduced from the data of (a) and (b) using a 10 $\mu$V criterion. In (d) we plot a conductance spectrum of a resistive bridge (C6 of Fig. 2 (c)) at 1.9 K and 0 T which shows a small asymmetry at low bias typical of magnetic tunneling junctions.      }
\end{figure}

Next we looked into the properties of the superconductive bridges of Fig. 2 (c). Fig. 3  depicts I-V curves (IVCs) at 2.1 K of the C7 bridge under various fields after ZFC (a), and after 1 T FC (b). The data looks very similar in both cases, as is even better seen in (c) where the critical current $\rm I_c$ deduced from this data using a 10 $\rm \mu$V criterion is plotted versus field. Since vortex physics and pinning should depend on the cooling conditions under different magnetic fields and under field cycling, one would expect hysteresis effects in the IVCs and different $\rm I_c$ values under ZFC and 1 T FC. This however is not the case, apparently due to strong pinning in the YBCO islands and hardly any vortex motion at 2.1 K, except for the data of Fig. 3 (b) at 0 T after 1 T FC and at 0 T after field cycling to 6 T and back to 0 T, where a vortex jump is observed at low positive bias. We have also found no hysteresis under field while measuring several IVCs of the C7 bridge at 2.1 K under increasing field up to 6 T and decreasing field back to 0 T (not shown). As no hysteresis effects were observed within the noise of the measurements, not under ZFC or FC conditions, neither under various fields and field cycling, we make the conjecture that will later be verified, that the magnetic properties of the trilayer are suppressed by the edge currents of the topological BS layer at the interface. Nevertheless, a weak signature of ferromagnetism is found in the resistive C6 bridge of Fig. 2 (c), whose conductance spectrum at 1.9 K and under 0 T is shown Fig. 4 (d). In this tunneling spectrum between SC islands, one can see a small asymmetry near zero bias which is typical of magnetic tunneling junctions \cite{TedrowMeservey,Aronov}. Thus the contribution of the ferromagnetic SRO layer to this spectrum is established.\\

\begin{figure} \hspace{-20mm}
\includegraphics[height=8cm,width=11cm]{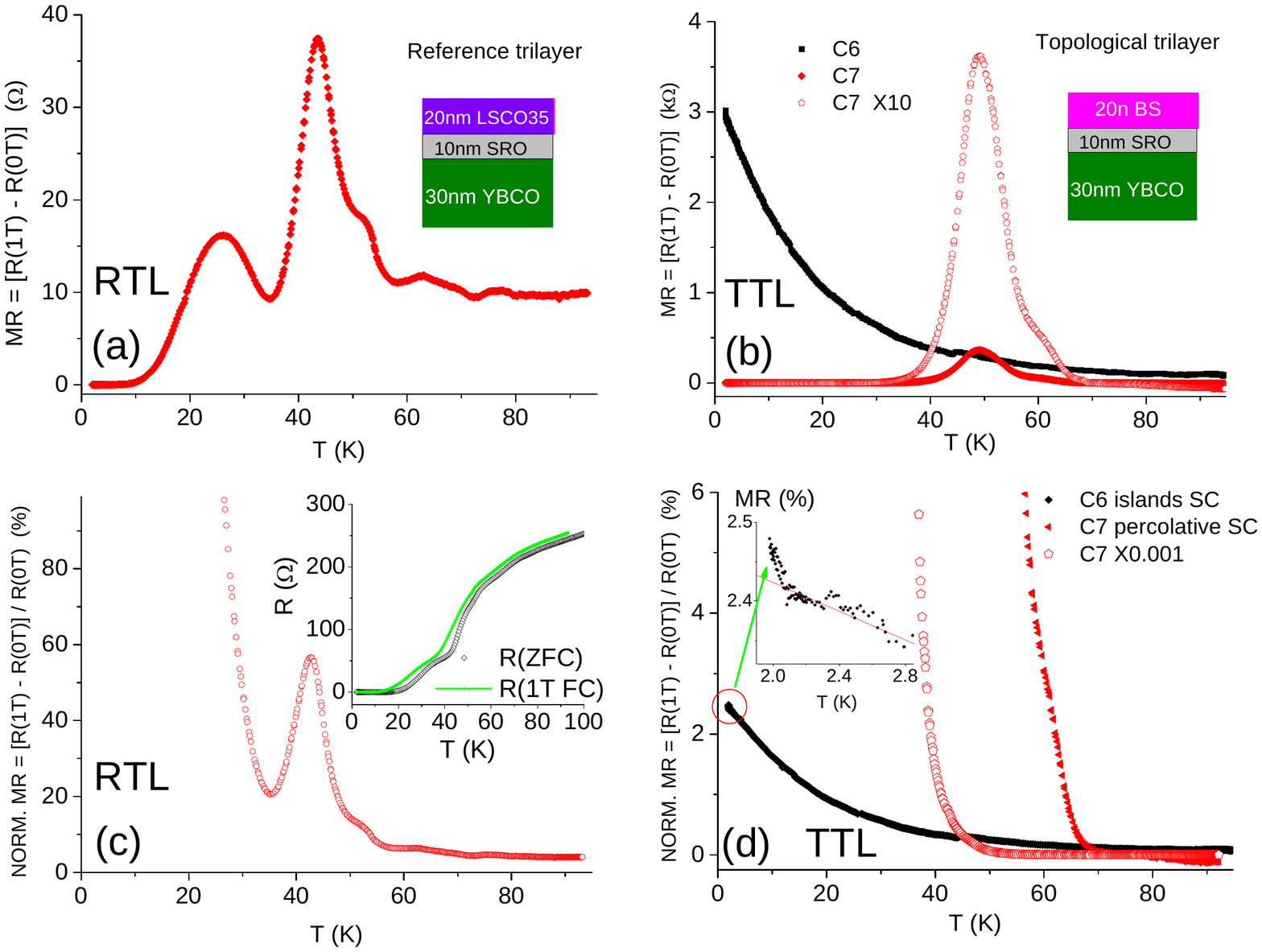}
\hspace{-20mm} \caption{\label{fig:epsart} Magnetoresistance vs temperature of $\rm 10\times 100\,\mu m^2$  microbridges of a RTL of 20 nm LSCO35 on 10 nm SRO on 30 nm YBCO is given in (a) and (c), and that of the the resistive (C6) and superconductive (C7) microbridges of the TTL of Fig. 2 (c) is shown in (b) and (d). The magnetoresistance is defined here as MR=R(1T)-R(0T) and its normalized value as [R(1T)-R(0T)]/R(0T). For the superconductive microbridges R(0T)$\rightarrow$0 at low temperature, thus their corresponding normalized MR diverges. The inset of (c) shows the R(ZFC) and R(1T FC) vs temperature of the RTL, and that of (d) is a zoom-in on the low temperature normalized MR of the resistive bridge C6 of Fig. 2 (c).      }
\end{figure}

To further explore the phenomena observed in the TTL of Fig. 2 (c) where resistive and superconductive bridges were found on the same wafer, we measured the magnetoresistance MR=[R(1T)-R(0T)] (in $\Omega$) and the normalized MR=[R(1T)-R(0T)]/R(0T) (in \%) of two such bridges vs temperature. The results are plotted in Fig. 4 (b) and (d), where we chose to present data of two adjacent bridges one resistive and insulating (C6) and the other metallic and superconducting (C7). For comparison we also show in Fig. 4 (a) and (c) MR results measured on one bridge of the RTL where a metallic LSCO35 layer replaces the topological BS layer in the trilayer. Since the MR data of the RTL is more easily interpreted and similar to the previously reported MR measurements on 5-50 nm LSCO35/5 nm LSCO12 \cite{KorenMillo}, we shall discuss it first. In Fig. 4 (a) of the RTL one sees two dominant peaks at 37 and 26 K, with leading edges on their high temperature side at 55 and 35 K, respectively. (The additional small peaks at higher temperatures originate in more oxygenated regions at the inner parts of the YBCO grains, and will not be discussed here.) The main peak originates in the superconducting transition of the YBCO grains, and the one on its left is a result of the second transition (see inset to Fig. 4 (c)) due to proximity effect of these YBCO grains with the cap layer of LSCO35/SRO. Both MR peaks are formed due to two competing phenomena, flux flow against flux pinning. On cooling down below $T_c$ at about 55 K, the MR is increasing rapidly due to increasing flux flow. By further lowering the temperature, pinning of flux also increases and as a result the MR decreases too, forming the main peak. In a stand-alone YBCO layer, this continues until there is no flux motion and the resistance goes to zero. Here however we have a trilayer and due to the proximity effect in it, there is a second superconductive transition at lower temperature as seen in the inset to Fig. 4 (c). Therefore, a second PE MR peak due to this transition is developed at low T, but in this case the resistance does go to zero below it. This effect is clearly seen in Fig. 4 (c) where the normalized MR diverges at low T. The main MR peak is still clearly seen, but from the second PE MR peak only the leading edge remains.\\

We now turn to the MR results of the TTL. Fig. 4 (b) shows a single MR peak for the superconductive C7 bridge of Fig. 2 (c), similar to the main MR peak of the RTL bridge in Fig. 4 (a). The second PE MR peak as in the RTL is now absent or within the noise of the measurements. This indicates that the PE in the TTL is strongly suppressed as compared to that of the RTL. Since the only difference between the TTL and the RTL is the top BS layer, we conclude that the PE suppression in the TTL is due to the apparently strong surface currents in the topological layer at the interface with the SRO. We know of no theory that predicts this strong PE suppression behavior with a topological material. The MR vs T of the resistive C6 bridge looks quite different from that of the superconductive C7 bridge, but actually it isn't. Its' slow and broad increasing MR below about the same 65 K, seems to be the leading edge of a very broad MR peak, just like an extended region of the C7 peak between 65 and 55 K. This indicates that transport in the C6 bridge occurs between superconducting islands which are coupled by the TI-FM cap layer. In Fig. 2 (c) however, only a shallow plateau is observed for C6 around 50-70 K with no signature of a superconductive transition. This is apparently due to the much higher resistance of the resistive bridges as compared to the superconductive ones above $\rm T_c$. A similar islands superconductivity was observed in bilayers of non-superconducting LSCO35 and superconducting underdoped $\rm La_{1.88}Sr_{0.12}CuO_4$ \cite{KorenMillo}. Thus to summarize, the conduction mechanisms in Fig. 2 (c) are such that C7 has a percolative SC while C6 has islands SC. The origin of the faster MR increase near 2 K in the inset of Fig. 4 (d), seems to be due to increasing slope of the MR peak. It can't be due to the Kondo effect, since it increases rather than decreases with increasing field.\\

\begin{figure} \hspace{-20mm}
\includegraphics[height=8cm,width=12cm]{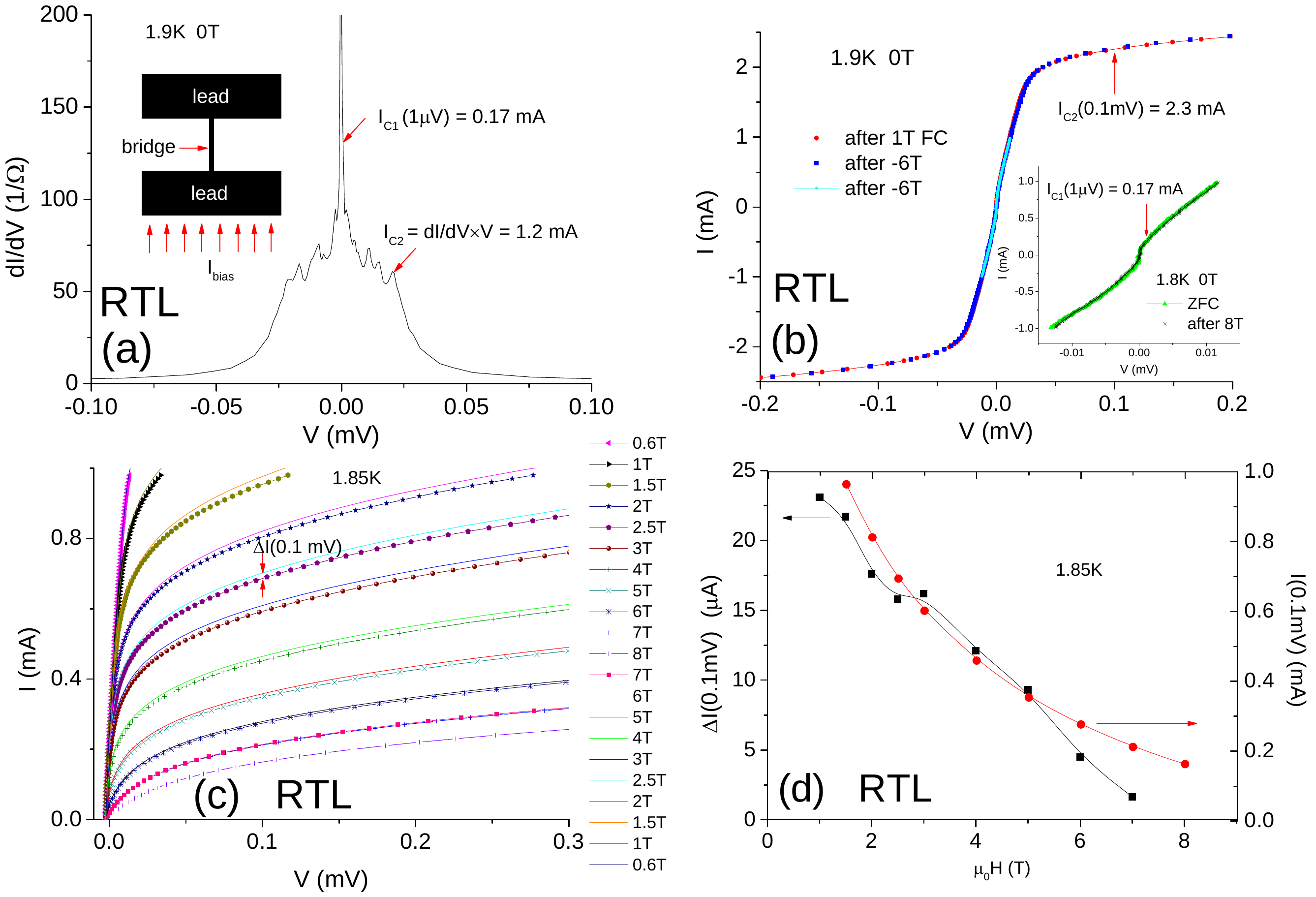}
\hspace{-20mm} \caption{\label{fig:epsart} (a) A conductance spectrum of a $\rm 10\times 100\,\mu m^2 $ bridge (see inset) of the RTL of Fig. 4 (a) and (c) (20 nm LSCO35 on 10 nm SRO on 30 nm underdoped YBCO). (b) depicts IVCs of this RTL bridge at 1.9 K and 0 T after various field cycling histories. No hysteresis is observed at 0 T. (c) shows many IVCs at 1.85 K under increasing and decreasing field, with hysteresis where larger currents are found at the same V-bias  under decreasing field. In (d) $\rm \Delta I$ and I at 0.1 mV of the IVCs in (c) are given vs field. The curves are just guides to the eye.     }
\end{figure}

In the following we study in more details the RTL properties with and without magnetic fields. Fig. 5 (a) depicts a conductance spectrum of the RTL bridge of Fig. 4 (a) and (c) at 0 T after ZFC, with a schematic drawing of the bridge and its leads in the inset. This spectrum shows two critical currents $\rm I_{c1}$ and $\rm I_{c2}$ which are attributed to the bridge and its leads, respectively. In Fig. 5 (b) a few IVCs are given at zero field again, but this time after various field cycling histories. The $\rm I_{c1}$ and $\rm I_{c2}$ are marked on these curves using different voltage criterions or methods of calculation than in (a). Clearly, for bias current above $\rm I_{c1}$, the bridge becomes resistive as seen by the slope at low bias in (b), but at higher bias the second $\rm I_{c2}$ of the leads is observed. The 0.1 mV criterion for this supercurrent was chosen arbitrarily for convenience. No hysteresis was observed in (a) and (b) under zero field. In Fig. 5 (c) several IVCs at 1.85 K are shown under increasing field up to 8 T followed by decreasing it back to 0 T. One can easily see that hysteresis is observed between 1 and 7 T after this field cycling. We use $\Delta$I(0.1 mV) as a measure of this hysteresis, and plot it vs field in Fig. 5 (d) together with the corresponding currents I under decreasing fields. The hysteresis clearly results from the ferromagnetic SRO layer of the RTL. Although this layer is present also in the TTL, no such hysteresis effects were observed in the TTL. This is a clear indication for strong suppression of the ferromagnetic properties by the edge currents of the topological layer, which is an effect similar to the previously observed suppression of the superconductive PE. \\

\begin{figure} \hspace{-20mm}
\includegraphics[height=8cm,width=10cm]{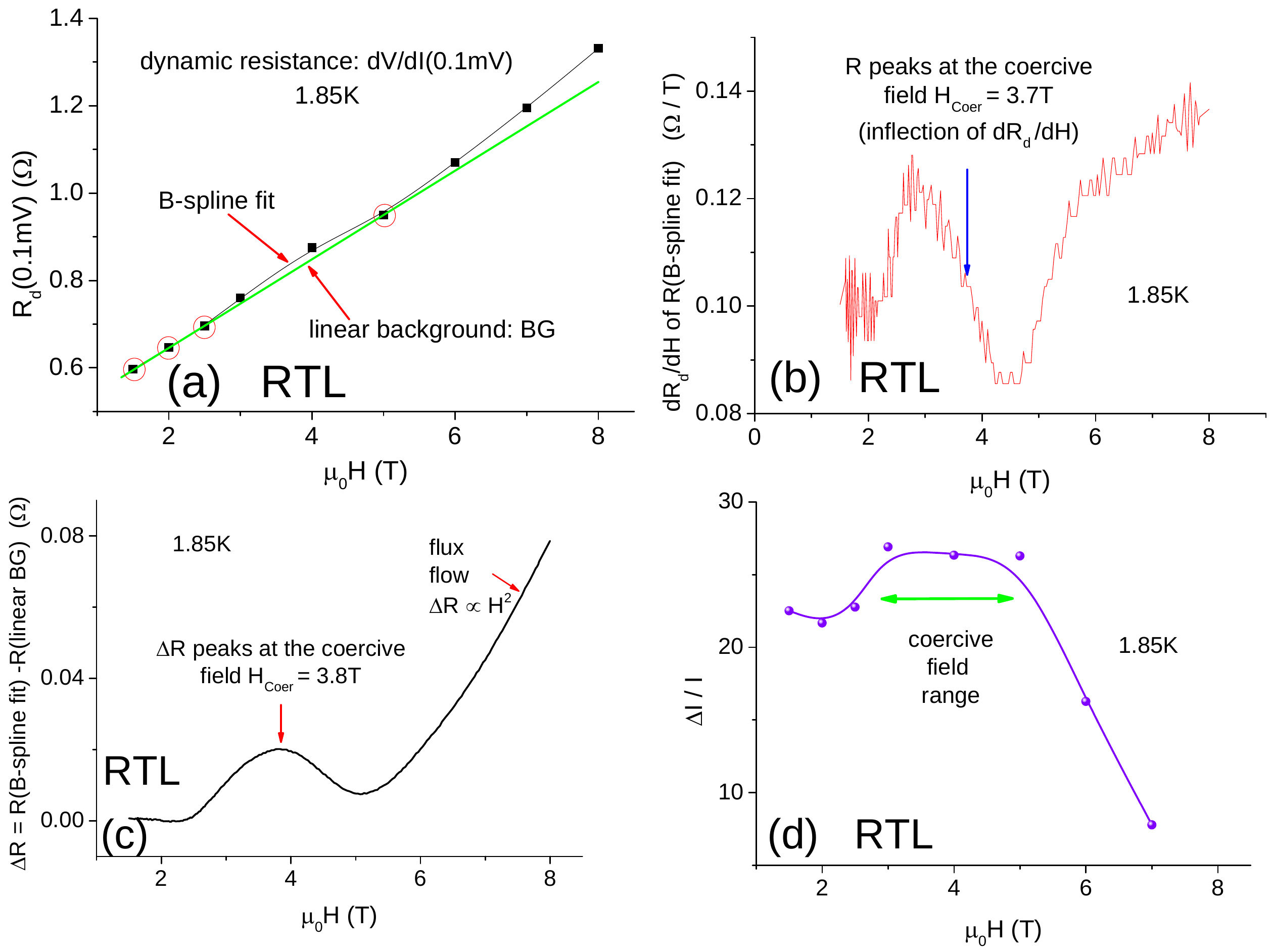}
\hspace{-20mm} \caption{\label{fig:epsart} Dynamic resistance $\rm R_d$ of the RTL of Fig. 4 (a) and (c) is shown at 1.85 K and 0.1 mV  in (a) versus magnetic field H. A linear fit back ground (BG, the green line) is obtained using the four encircled data points, and the B-spline fit is calculated using all data points. (b) shows the field derivative of the B-spline fit of (a), while (c) depicts  $\rm \Delta R$ which is the difference between  R of the B-spline fit of (a) and the linear background BG at the same H. (d) shows $\rm \Delta I/I$ using the data of Fig. 5 (d) vs field.       }
\end{figure}

To further explore the hysteresis effect of the RTL, we now look for a signature of the coercive field ($\rm H_{Coer}$) of the SRO in this trilayer. Previously, we found $\rm H_{Coer}$ of about 1 T in similar 15 nm BST/15 nm SRO bilayers \cite{KorenMPE}. Here the SRO layer is thinner, only 10 nm thick, thus  a higher $\rm H_{Coer}$ is expected. Fig. 6 (a) shows the dynamic resistance $\rm R_d$=dV/dI at 1.85 K derived from the IVCs of Fig. 5 (c) at 0.1 mV. A B-spline fit to the data and a linear fit through a selected set of data points are also depicted in this figure. To identify a peak in the data of (a) at $\rm H_{Coer}$ we first take the field derivative of the B-spline fit. The result is plotted in (b) where the inflection point indicates the field where $\rm R_d$ peaks which leads to $\rm H_{Coer}$ of 3.7 T. Alternatively, if we consider the linear fit in (a) as background (BG), we can subtract it from the B-spline fit and the result is depicted vs field in (c). Here we can clearly see the coercive field peak at 3.8 T which is almost equal to that obtained from (b) to within the noise of the derivative curve. The parabolic behavior of $\rm \Delta$R at high fields in (c) is common, and due to weak flux flow that still remains at this low temperature \cite{Tinkham}. $\rm H_{Coer}$ value around 4 T can also be found from a plot of $\rm \Delta$I/I (calculated from the data of Fig. 5 (d)) vs H which is given in Fig. 6 (d). This data is quite scattered and therefore only a range of maximum $\rm \Delta$I/I can be obtained between 3 and 5 T, which agrees with the values found earlier. We conclude from Figs. 5 and 6 that the ferromagnetic properties of the SRO layer in the RTL are robust, while in the TTL they are strongly suppressed as in Fig. 3. We attribute this suppression, as well as that of the superconductive PE MR of Fig. 4 (b) and (d), to the strong edge currents in the topological BS layer of the present trilayers. \\

\begin{figure} \hspace{-20mm}
\includegraphics[height=8cm,width=10cm]{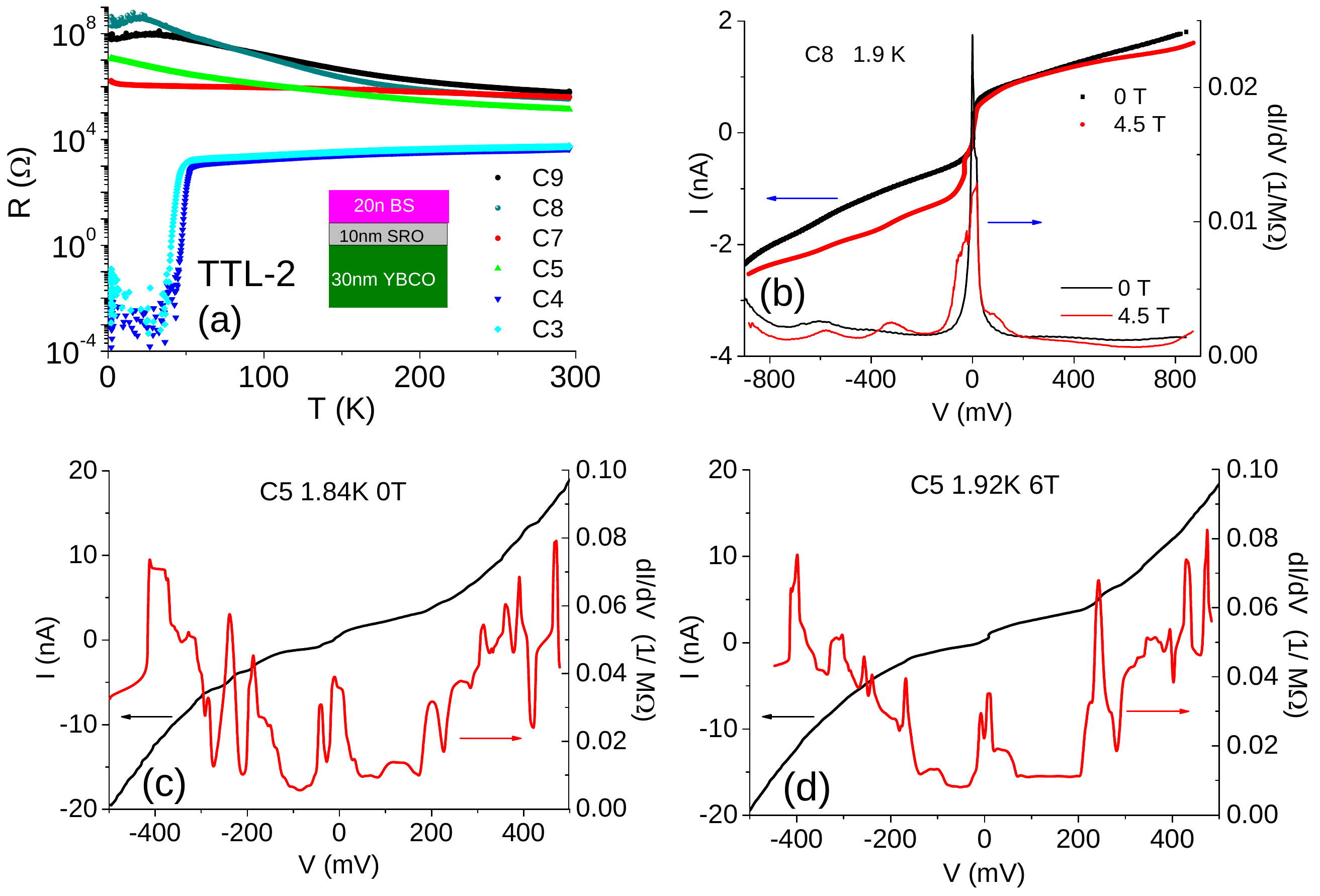}
\hspace{-20mm} \caption{\label{fig:epsart} (a) R vs T of six microbridges on a different topological trilayer TTL-2 whose resistive bridges have one to three order of magnitude higher resistance than that of our previous TTL of Fig. 2 (c). IVCs and conductance spectra of the C8 and C5 bridges at about 1.9 K are depicted in (b), (c) and (d). In (b) the magnetic fields are 0 and 4.5 T, while in (c) and (d) they are 0 and 6 T, respectively.        }
\end{figure}

Next we look for signatures of Majorana fermions in our data. For this we chose to present data measured on a different topological trilayer TTL-2, whose resistive bridges have one to three orders of magnitude higher resistance than that of our previous TTL of Figs. 2 (c). The results are shown in Fig. 7 where (a) depicts R vs T of six bridges on this wafer. Compared to the resistance of the C6 bridge of Fig. 2 (c) at about 2 K, the C8 bridge here has about thousand times higher resistance while the C5 bridge is ten times more resistive. Fig. 7 (b), (c) and (d) show IVCs and conductance spectra of the two bridges at about 1.9 K and under various fields. Fig. 7 (b) shows the data for the most resistive C8 bridge, which depicts a single prominent zero bias conductance peak (ZBCP) at 0 T that is strongly suppressed by field.  Also seen are a few much weaker bound state peaks and knees whose V-values are field dependent. The data of the two orders of magnitude less resistive C5 bridge of (c) and (d) is not as clear cut. While the tunneling nature of these spectra is clearly seen and both have ZBCPs, many additional peaks also appear asymmetrically at different V-values which depend on the applied field. We note that even under 0 T, an internal field exists in the trilayer due to the magnetization of the SRO layer. To qualitatively understand these results, we refer to the model in Fig. 1 (d), and in particular to the area enclosed by the dashed rectangle. As already discussed earlier, our trilayers with underdoped YBCO form a network of isolated YBCO grains, thus current can't flow directly between these grains but can flow via the BS/SRO cap layer. The edge currents in the topological BS layer are marked schematically in Fig. 1 (d) by the two bold black arrows. It seems that these currents are responsible for the strongly suppressed superconductive PE and ferromagnetism as discussed previously. Therefore, it is highly probable that the edge current closer to the interface induces a proximity edge current in the adjacent SRO/YBCO layers (see the thin black arrow in Fig. 1 (d)). In this case, one can use the model given by Beenakker \cite{Beenakker} and conclude that a spatially split pair of Majorana fermions exists (the red dots in Fig. 1 (d)) which is bound by the magnetic Zeeman gap and the superconducting gap on both sides, thus leading to a Majorana bound state (MBS) at zero bias in the present system. This bound state in turn will increase the density of states at zero bias, which will lead to a ZBCP in the conductance spectra. ZBCPs in general can also originate in Andreev bound states (ABSs) \cite{Tanaka,Aronov}, but the presently observed ZBCPs in Fig. 7 (b)-(d) are consistent with the MBS scenario just as well. In a network of isolated superconducting grains as in Fig. 1 (d), there can be many MBSs which also interfere with one another. Thus the many other peaks beside the ZBCPs in (c) and (d) can be a result of such interference or due to non zero bias ABSs. The derivatives dI/dV in Fig. 7 (c) and (d) though are quite noisy, and some of the peaks could  originate in noise. Therefore, it is even better to look at the original IVCs, all of which have a step at zero bias, that leads to the ZBCPs. We attribute these steps to MBSs, but they could also result from residual small supercurrent inside the YBCO grains. Thus we conclude that the observed ZBCPs could originate in MBSs, but ABSs and small supercurrents could not be ruled out as their source. Moreover, for the bridge with the highest resistance (C8), a smaller number of weaker peaks are found in the conductance spectra besides the ZBCP, as compared to the C5 results. This is very reasonable considering that in the 1D network of channels (the weak-links) via which the current flows between the YBCO grains, a higher resistance means that less channels are open for this flow. Thus fewer channels give rise to a robust ZBCP with very little interference peaks, as we actually observed. \\

Finally, we discuss the potential of the present trilayers with their 1D network of weak-link channels, as a possible platform for Majorana electronics. We compare our results to the recent work by the Marcus group \cite{Marcus}, which also reports on a 1D artificial network of SC/TI wires grown selectively on a wafer, without an FM layer but with gating. Both systems could be useful for Majorana electronics since they facilitate braiding of MBSs necessary for quantum computing. But while our 1D network is disordered and hard to control at the present time, theirs is ordered, have good control and is much more advanced. Nevertheless, the operational temperatures of their devices are at the tens of milii Kelvins, while ours can be easily operated at a few Kelvins due to the much higher $\rm T_c$ of YBCO (50 K here) vs Al (1.2 K). Possibly, a combination of the two techniques, namely selective growth or patterning of the wires in an ordered network, and the use of a superconductor with a higher $\rm T_c$ instead of Al, would yield a more acceptable and controllable system for practical applications. We conclude with a few remarks on our TTLs. As seen in Fig. 7 (b), it is desirable to use highly resistive bridges for producing MBSs. It is also expected that patterning of much smaller nano-bridges in the present TTLs would allow to isolate a smaller number of MBSs, thus reducing the number of interference peaks. Moreover, the use of other topological materials such as semiconductors with  strong spin orbit interaction and good lattice match with the base SRO/YBCO bilayer, instead of the presently used BS, is expected to lead to all-epitaxial trilayers with much better control of the MBSs properties. This could open the way for the use of such TTLs as a possible  platform for Majorana based nano-electronics on standard wafers. \\

\section{Conclusions}

To summarize, microbridges patterned in the TTL  of the underdoped 60 K YBCO phase showed either percolative superconducting behavior or resistive tunneling conductance between YBCO islands via the BS/SRO cap layer.  Signatures of the comprising BS and SRO layers of the TTL were found, where WAL in the magnetoresistance of the BS layer and asymmetry in the tunneling conductance spectrum of the SRO layer were observed. Compared to the RTL results, we find a strong suppression of the superconductive PE as well as the ferromagnetic properties in the TTL. This is attributed to proximity induced edge currents in the SRO/YBCO bilayer at the interface by the topological layer. Observed ZBCPs in our conductance spectra in highly resistive TTL brides is consistent with the presence of MBSs.  \\


\bibliography{AndDepBib.bib}

\bibliography{apssamp}

\end{document}